\documentclass[useAMS,usenatbib]{mn2e}

\usepackage{amssymb,amsmath}
\usepackage{lscape}
\usepackage{graphicx}
\usepackage{lscape}

\newcommand{\Min}{_{\mathrm{min}}}

\newcommand{\Sun}{_{\sun}}

\newcommand{\Feedb}{_\textrm{fb}}
\newcommand{\SF}{_\textrm{sf}}
\newcommand{\FF}{_\textrm{ff}}
\newcommand{\Cooloff}{_\textrm{coff}}
\newcommand{\Delay}{_\textrm{delay}}
\newcommand{\MFB}{_\textrm{mfb}}
\newcommand{\EFB}{_\textrm{efb}}

\newcommand{\K}{\,\textrm{K}}

\newcommand{\Kpc}{\,\textrm{kpc}}

\newcommand{\Yr}{\,\textrm{yr}}

\newcommand{\Myr}{\,\textrm{Myr}}

\newcommand{\Kms}{\,\textrm{km}\,\textrm{s}^{-1}}

\newcommand{\gccm}{\,\textrm{g}\,\textrm{cm}^{-3}}

\title[SF in shocked cluster spirals]{Star formation in shocked cluster spirals and their tails}
\author[Roediger et al.]{
E.~Roediger$^{1}$\thanks{E-mail:
eroediger@hs.uni-hamburg.de (ER)},
M.~Br\"uggen$^{1}$,
M.~S.~Owers$^{2}$,
H.~Ebeling$^{3}$,
M.~Sun$^{4}$
\\
$^{1}$Hamburger Sternwarte, Universit\"at Hamburg, Gojenbergsweg 112, D-21029 Hamburg, Germany\\
$^{2}$Australian Astronomical Observatory, PO Box 915, North Ryde, NSW 1670, Australia\\
$^{3}$Institute for Astronomy, University of Hawaii, 2680 Woodlawn Drive, Honolulu, HI 96822, USA\\
$^{4}$Department of Physics, University of Alabama in Huntsville, Huntsville, AL 35899, USA
}
\begin{document}

\date{Accepted 1988 December 15. Received 1988 December 14; in original form 1988 October 11}

\pagerange{\pageref{firstpage}--\pageref{lastpage}} \pubyear{2011}

\maketitle

\label{firstpage}

\begin{abstract}
Recent observations of ram pressure stripped spiral  galaxies in clusters revealed details of the stripping process, i.e., the truncation of all interstellar medium (ISM) phases and of star formation (SF) in the disk, and multiphase star-forming tails. Some stripped galaxies, in particular in merging clusters, develop spectacular star-forming tails, giving them a jellyfish-like appearance. 
In merging clusters, merger shocks in the intra-cluster medium (ICM)  are thought to have overrun these galaxies, enhancing the ambient ICM pressure and thus triggering SF, gas stripping and tail formation. 
We present idealised hydrodynamical simulations of this scenario, including standard descriptions for SF and stellar feedback. To aid the interpretation of recent and upcoming observations, we focus on particular structures and dynamics in SF patterns in the remaining gas disk and in the near tails, which are easiest to observe. 
The observed jellyfish morphology is qualitatively reproduced for, both, face-on and edge-on stripping. In edge-on stripping, the interplay between the ICM wind and the disk rotation leads to asymmetries along the ICM wind direction and perpendicular to it. The apparent tail is still part of a highly deformed gaseous and young stellar disk. In both geometries, SF takes place in knots throughout the tail, such that the stars in the tails show no ordered age gradients. Significant SF enhancement in the disk occurs only at radii where the gas will be stripped in due course.
\end{abstract}


\section{Intro}  \label{sec:intro}
%
Spiral  galaxies in clusters have long been known to lose their gas via ram pressure stripping (RPS) due to their motion through the ICM (\citealt{Gunn1972}). The cluster infall and hence increase in ram pressure proceeds on time scales of several 100 Myr; thus, in general, RPS gradually removes a spiral's gas disk (e.g., \citealt{Roediger2007}) and quenches its star formation (SF) in an outside-in fashion. This overall scenario has been confirmed with numerical simulations by different groups and is observed for various spirals in, e.g., the Virgo cluster (reviews by \citealt{Roediger2009}, \citealt{Vollmer2013}, respectively). Recent observations revealed a wealth of structure in the remaining multiphase star-forming ISM disk and the stripped tails, including kinematic information and age structures for the stellar tails (e.g., \citealt{Abramson2014,Kenney2014,Jachym2014}, among others). A characteristic feature of the stellar tails are filaments and knots of young stars.

Several apparently extreme cases of  stripping have been observed in very massive clusters (\citealt{Owers2012}, O12 hereafter; \citealt{Ebeling2014}, E14 hereafter).  These galaxies show tails with bright blue knots 10 kpc and more downstream of the disks, indicating substantial SF in their tails, and asymmetric disks of young stars.  Since some of the host clusters of these systems are highly disturbed, we here focus on the impact of cluster mergers on the stripping of spiral galaxies. In merging clusters, galaxies can encounter  higher-velocity ICM head winds after being overrun by the merger shocks. The accompanying enhanced ICM pressure behind the shock could potentially boost the SF in these stripped galaxies (\citealt{Bekki2010sync}). 

We present idealised simulations of a spiral galaxy with a multi-phase ISM, self-regulated by SF and feedback, being overrun by an ICM shock followed by a fast ICM wind.
In contrast to previous work, we do not focus on the global evolution of the galaxy but follow the spatial and temporal evolution of SF in the model galaxy and its tail in face-on and edge-on stripping geometries.

\section{Method}  \label{sec:method}

Our simulations use the hydrodynamics grid code FLASH (version 4.0.1, \citealt{Fryxell2000flash}). Both ISM and ICM are treated as ideal gases. The ISM is set up in a rotating disk  in an analytic galaxy potential.  SF is modelled via turning gas into massive star particles, which move according to the local gravity, and serve as time-dependent sources of mass and energy for the gas (stellar feedback). Our simulations include self-gravity of the gas, gravity due to the newly formed star particles, and analytic potentials for the galaxy's dark matter halo, bulge, and old stellar disk.

The model galaxy is a massive spiral galaxy with a flat rotation curve at $200\Kms$ as  in \citet{Roediger2006}, with an old stellar disk of $10^{11}M\Sun$, a bulge of $10^{10}M\Sun$, a dark matter halo of $10^{11}M\Sun$ within 23 kpc, and a gas disk of $1.5\times 10^{10}M\Sun$. Initially, the model galaxy evolves 400 Myr in isolation (ambient gas temperature and density of $10^7\K$ and $0.5\times 10^{-27}\gccm$, respectively) until a  global equilibrium state of SF and feedback is reached.     The low ICM density and pressure environment minimizes the impact of the thermal ICM pressure on the ISM, but already truncates the ISM disk to a radius of 15 kpc. 

We then initialise a piston shock in the ICM that overruns the galaxy and is followed by a constant ICM wind of $2000\Kms$. The wind density ($1.2 \times 10^{-27}\gccm$) and temperature $11\times 10^7\K$) follow from the Rankine-Hugoniot shock conditions w.r.t. the previously static, low-temperature ICM. The ambient pressure is enhanced by a factor of 40. This ICM wind strips the galaxy substantially, but not completely. Its lower density but faster velocity and high temperature mimic a merger shock outside of a cluster core. The wind can hit the galaxy face-on or edge-on.

The generous size of our simulation box of $(-100\Kpc,$ $100\Kpc)^3$ ensures a free ICM  flow around the galaxy.  All sensible refinement criteria would quickly fully resolve the galactic disk as well as the tail of stripped gas. We therefore use a fixed nested grid and fully refine the disk region (radius 20 kpc, thickness 6 kpc), as well as the tail up to 30 kpc downstream of the galaxy centre. We use resolutions of 0.2 and 0.1 kpc for the face-on and edge-on runs, respectively. Resolution affects the strength of the initial star burst as discussed below, but the quasi-equilibrium states before and after the shock passage are independent of resolution.

\begin{table}
\caption{ISM model parameters.}
\begin{tabular}{lll}
\hline
SF density threshold        & $\rho\SF$       & $3\times 10^{-24}\gccm$  \\
SF temperature threshold & $T\SF$           & $1.5\times 10^{4}\K$ \\
SF efficiency parameter   & $\epsilon\SF$ & 0.2 (0.8 at $\Delta x=0.2\Kpc$) \\
feedback delay timescale & $\tau\Delay$   & 6 Myr\\
feedback timescale          & $\tau\Feedb$    & 8 Myr \\
mass feedback efficiency & $\epsilon\MFB$ & 0.1 \\
energy feedback efficiency & $\epsilon\EFB$ & $10^{-5}$ \\
\hline\hline
\end{tabular}
\label{tab:ISM}
\end{table}%
%
The simulations include radiative cooling from an optically thin plasma making use of the tables compiled by \citet{Wiersma2009} from the code CLOUDY (\citealt{Ferland1998}) with a cooling floor at  $ 10^4$K. 

SF and stellar feedback is modelled in the standard approach similar to \citet{Tonnesen2012} (TB12) and references therein. In short, SF occurs if the gas density exceeds a threshold density $\rho\SF$ and the gas temperature falls below a threshold temperature $T\SF$ (see Table~\ref{tab:ISM}). In such grid cells, gas is turned into a star particle of mass $m_p$ according to
\begin{equation}
m_p = \epsilon\SF\, \rho \, \Delta x^3 \, \Delta t / \tau\FF,  \label{eq:sf}
\end{equation}
where $\tau\FF$ is the local free-fall time, $\rho$ is the local gas density, $\Delta x^3$ the volume of the grid cell, $\Delta t$ the current timestep, and $\epsilon\SF$ the SF efficiency per free-fall time. We require a minimum particle mass of $m_p{}\Min=3\times 10^3 M\Sun$ to prevent slowing down the simulations and to match our model assumption that each star particle represents a whole stellar population. Due to the cooling floor the gas densities rarely reach values that would lead to the formation of star particles above the required mass minimum. If this condition is the only one that prevents the particle from being created, the particle is still  created with a probability equal to the ratio of predicted mass  (Eqn.~\ref{eq:sf}) to $m_p{}\Min$. If the particle passes this probability test, it is created with $m_p{}\Min$, which ensures the desired global SF efficiency is achieved (TB12). In this prescription, SF traces well the densest gas regions. Particles inherit the local gas velocity.

On formation, star particles represent giant molecular clouds rather than stars. To account for the evolution of the molecular cloud and the subsequent evolution of the stellar population inside the particle, we delay feedback activity of new star particles by $\tau\Delay$. After this delay time,  star particles inject mass and energy into their nearest grid cell over a timescale  $\tau\Feedb$ at a rate proportional to their mass:
\begin{eqnarray}
\dot m &=& \epsilon\MFB\, m_p(t) / \tau\Feedb\;\;\textrm{and}  \label{eq:mfb}\\
\dot E &=& \epsilon\EFB\, m_p(t) c^2 / \tau\Feedb,  \label{eq:efb}
\end{eqnarray}
where $\epsilon\MFB$ and $\epsilon\EFB$ are the feedback efficiencies for mass and energy, respectively. They account for a stellar mass loss of 10\% from a stellar population (\citealt{Agertz2011}) and one supernova of $10^{51}$ erg per $55M\Sun$ of formed stars (TB12). The particle mass is reduced accordingly, and the injected gas inherits the velocity from the star particle.  This feedback description is a simplified model of the temporal evolution of feedback from an ageing stellar population. Its simplicity, however, allows us to assess the interactions between ISM-internal timescales and timescales on which the galaxy's environment changes. A systematic investigation will be the focus of a companion paper. 

In order to account for feedback  on unresolved scales, cooling is switched off in grid cells that are receiving feedback currently or have received feedback within the previous time span of length $\tau\Cooloff=5\Myr$. This is done via a count-down tracer that is advected with the gas. This allows  feedback to inflate supernova bubbles instead of losing the energy instantaneously by radiative cooling in the high density gas cells (\citealt{Agertz2011}).

\section{Results}  \label{sec:shock}

\begin{figure*}
\includegraphics[trim=0 340 90 100,clip,width=0.68\textwidth]{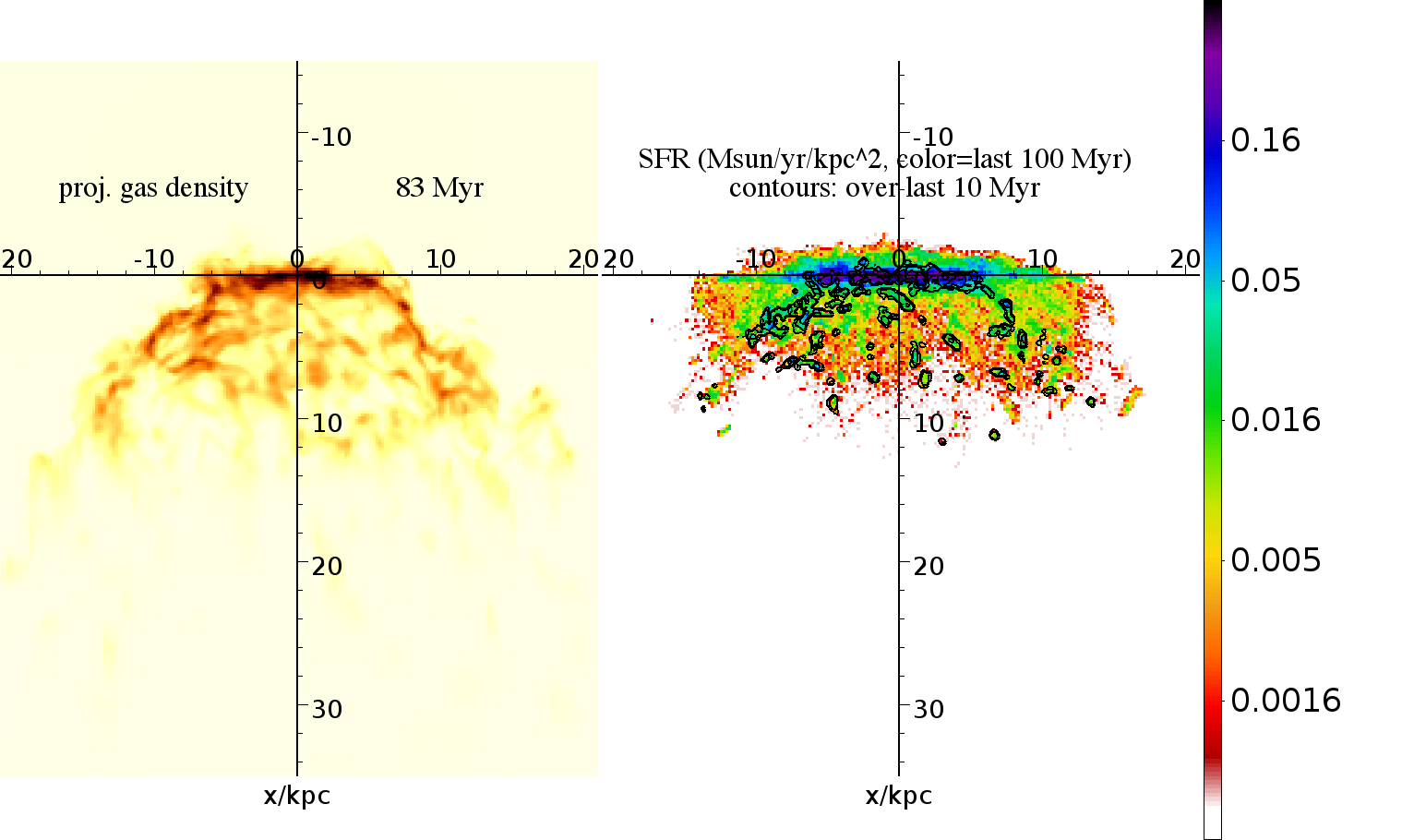}
\includegraphics[trim=300 1670 300 360,clip,width=0.3\textwidth,type=png,ext=.png,read=.png]{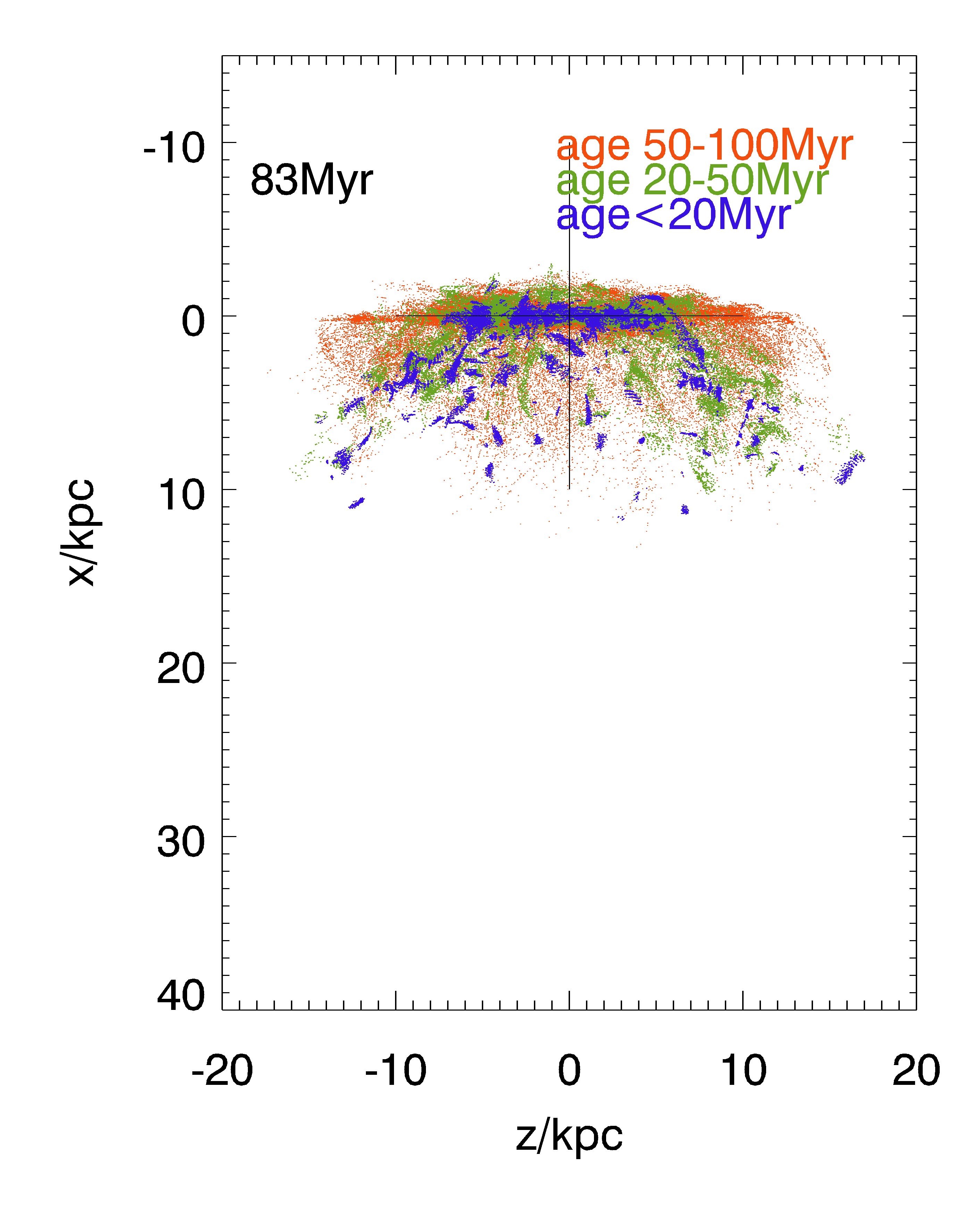}
\includegraphics[trim=0 170 90 100,clip,width=0.68\textwidth]{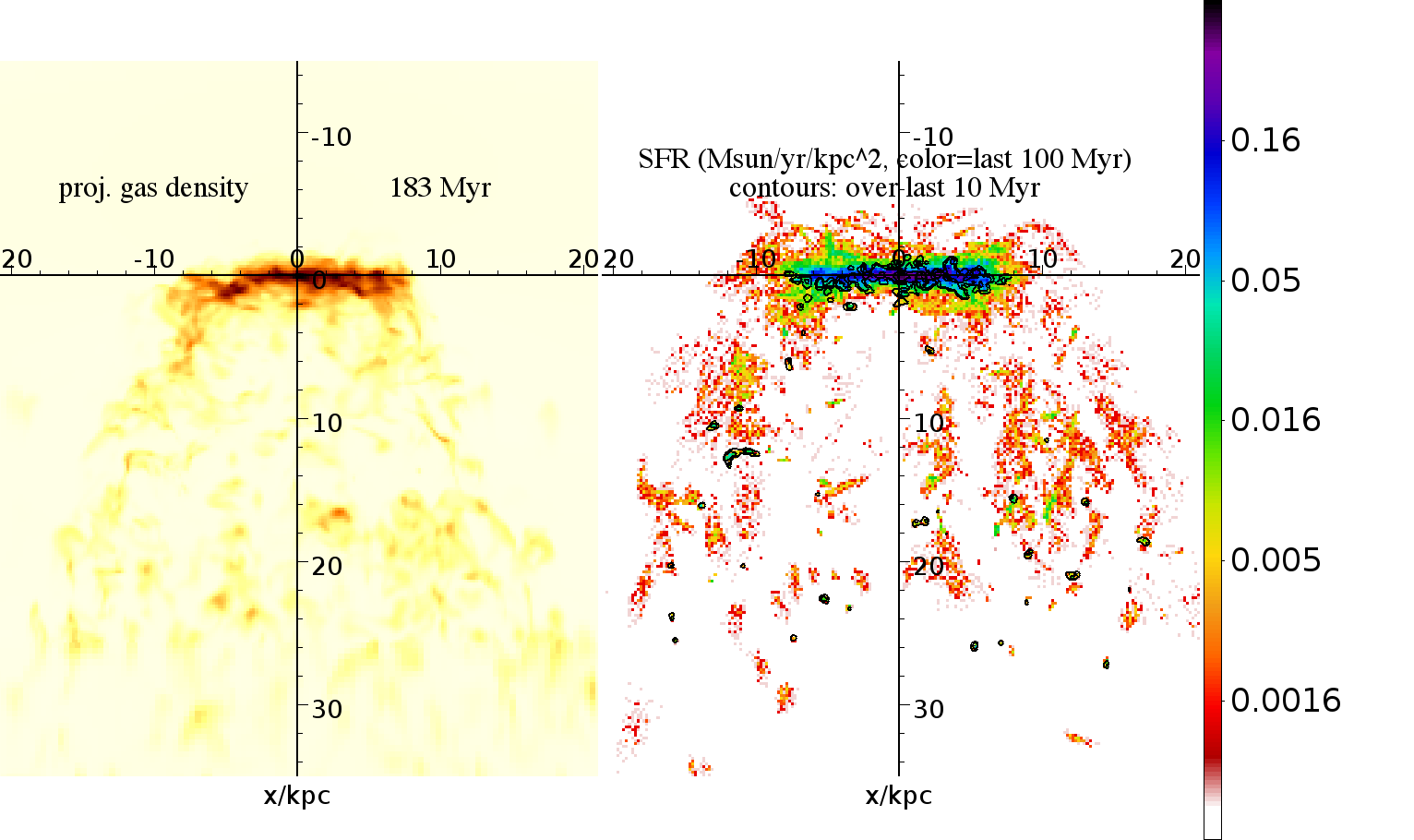}
\includegraphics[trim=300 1100 300 360,clip,width=0.3\textwidth,type=png,ext=.png,read=.png]{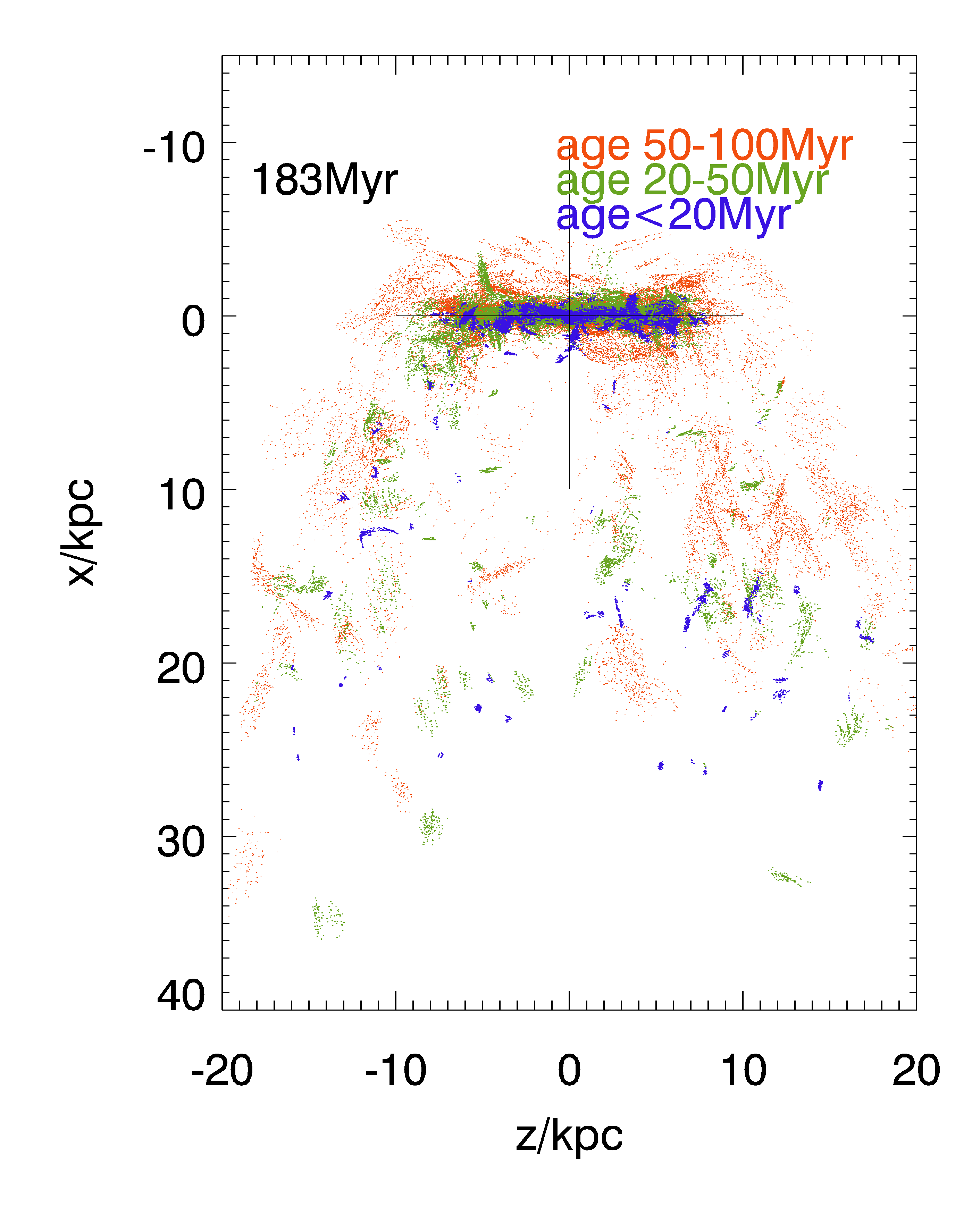}
\caption{Face-on shock passage and subsequent ICM wind from top to bottom. Projected gas density (left), projected SFR (middle), and ages of star particles (right) for 2 timesteps after shock passage (see labels in left column). Axes are labelled in kpc.  SF occurs in the remaining disk and throughout the tail. Averaging the recent SFR over the last 100 Myr (colour-coded in blue) and the last 10 Myrs (contour levels $0.003, 0.03, 0.3$) results in different patterns as discussed in the text. Animations can be found at 
\url{http://www.hs.uni-hamburg.de/DE/Ins/Per/Roediger/research.html\#Spiral}.}
\label{fig:sf_on_gas_faceon2000}
\end{figure*}

The gas removal in the face-on stripping proceeds in the same manner as described in the literature (e.g., \citet{Kapferer2009,Tonnesen2011}, see Fig.~\ref{fig:sf_on_gas_faceon2000}). After the shock passes the galaxy the ICM wind continues to push the galaxy gas downstream. Outside of  the stripping radius the ICM ram pressure can overcome the galaxy's gravity and push out the disk gas. It takes several 10 Myr to move the gas out of the galaxy, i.e., the actual gas removal is somewhat delayed. At the same time, SF takes place continuously in the unstripped and stripped gas. Thus,  at later times, observations of SF tracers that average over more than the recent $\sim 10\Myr$ will show SF at already gas-free radii in the disk.  

The stripped gas continues to form stars in knots and filaments throughout the tail. After $\sim 80\Myr$ (130 Myr) a $\sim 10\Kpc$ (20 kpc) long star-forming tail is established. The  SFR averaged over the last 10 Myr in the complete tail ($>$ 4 kpc away from the disk plane) varies between about 0.4 and and $1.5 M\Sun \Yr^{-1}$. Averaging the observed 'recent' SF over longer times averages out the actual filaments and knots of  SF.  In this instantaneous stripping scenario, SF in the tail does not proceed in the systematic 'fireball' scenario described in \citet{Kenney2014}, where stripped and accelerated clouds are thought to continuously form stars, which drop out of the accelerated gas cloud and form trails of stars falling back to the disk. In this fireball scenario, the  star filaments in the tail should have clear age gradients. In contrast, we find SF in knots throughout the tail. Short stellar filaments arise due to the backfall dynamics of the unbound single-age star clusters, or appear to arise due to favorable projection. In agreement with TB12, most  stars in the tail do not reach escape velocity and fall  back through the disk,  forming upstream filaments of moderately recent SF several kpc above the disk (see also \citealt{Kapferer2009}). Tracers of very recent SF ($\lesssim 20\Myr$) do not show this effect, as truly upstream SF is rare.

\begin{figure*}
\includegraphics[trim=0 120 40 85,clip,width=0.95\textwidth]{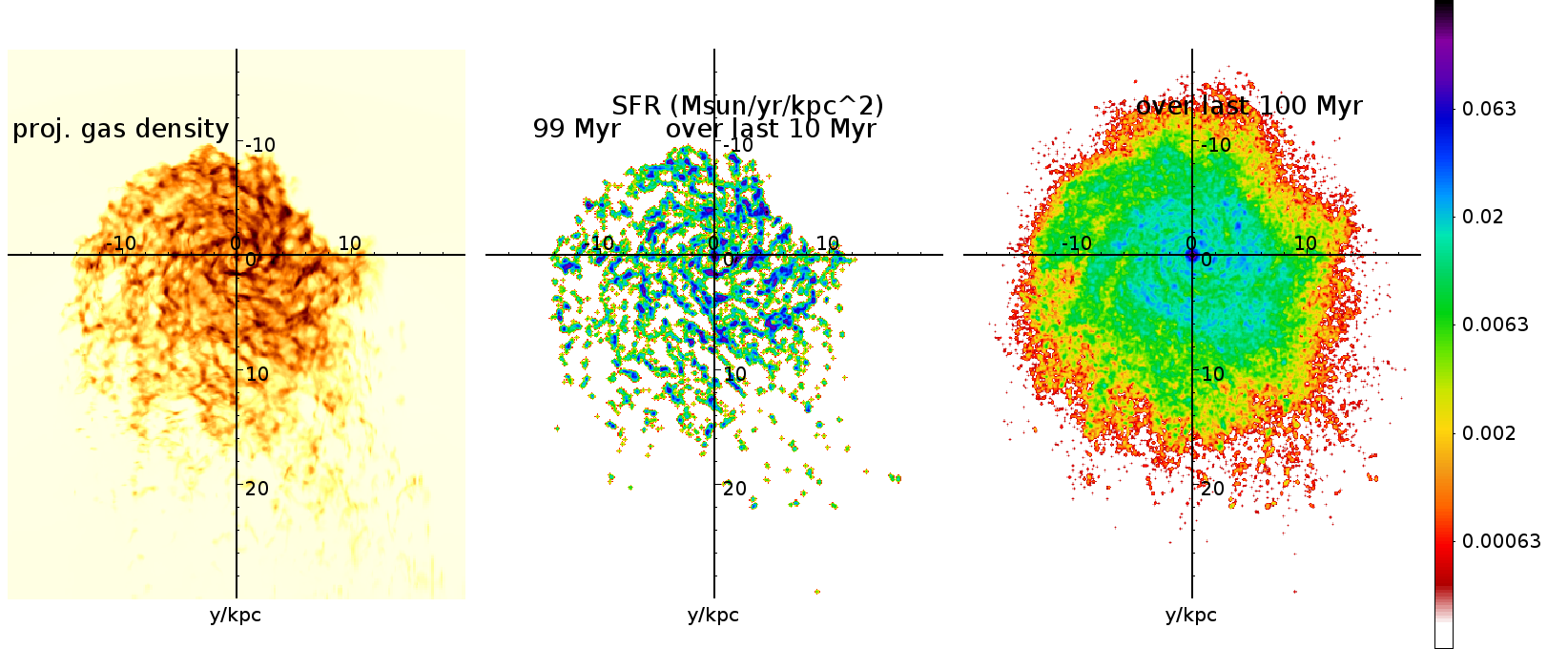}
\includegraphics[trim=0 50 40 85,clip,width=0.95\textwidth]{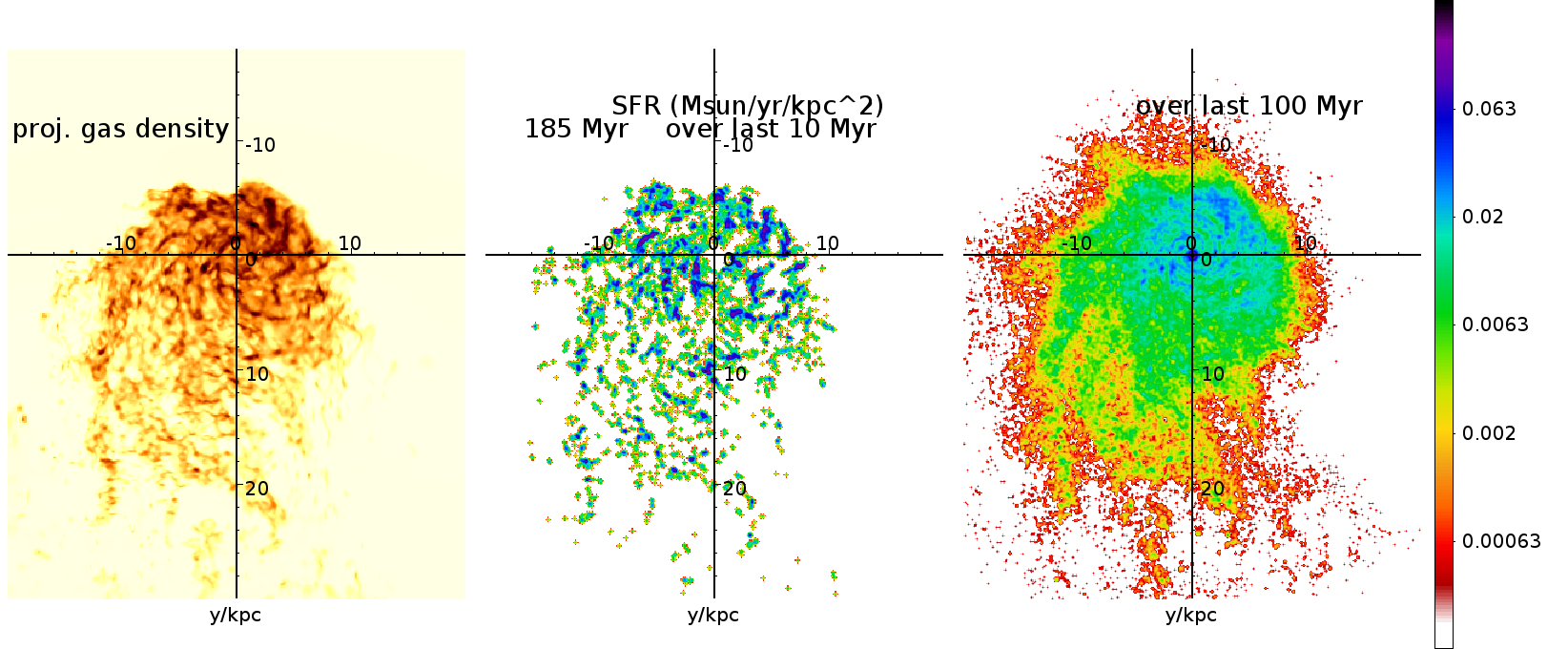}
\includegraphics[trim=0 20 40 85,clip,width=0.95\textwidth]{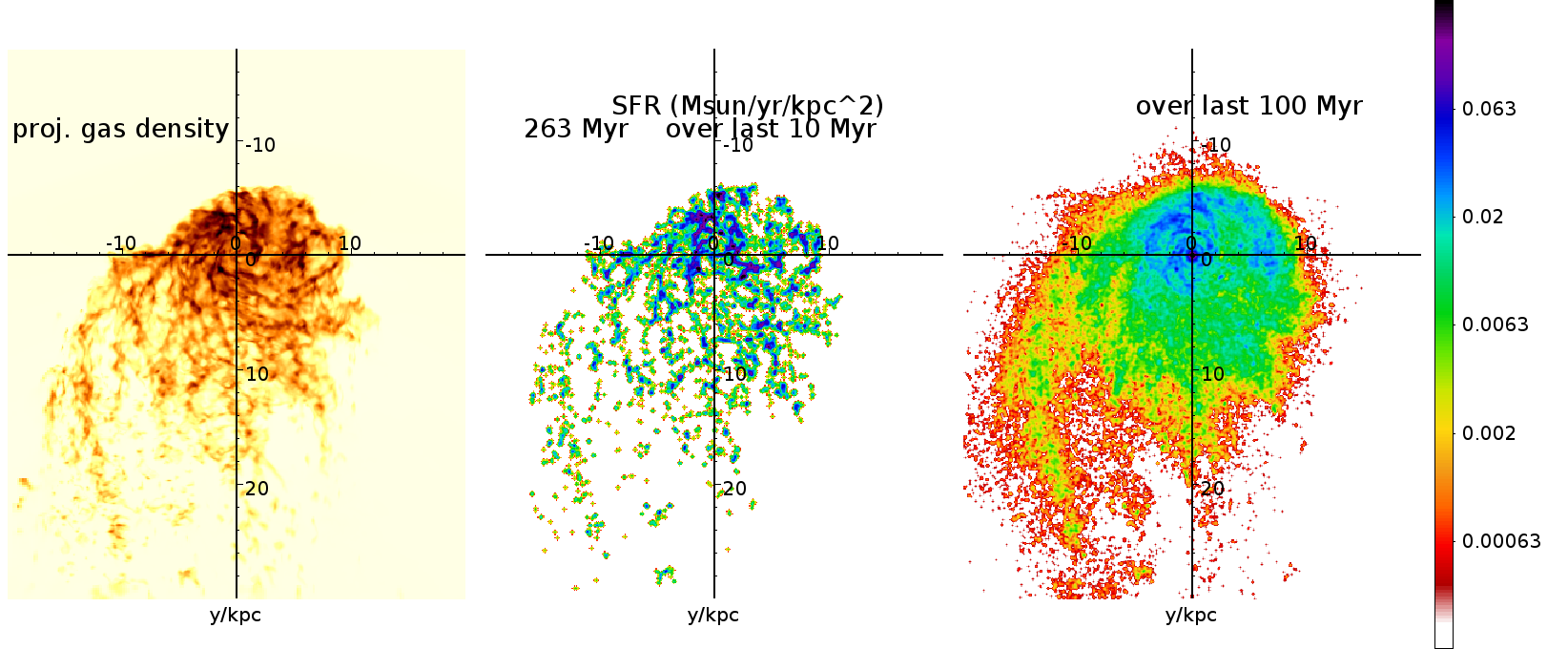}
\caption{Edge-on shock passage and subsequent ICM wind from top to bottom, one timestep per row. The time since shock passage is indicated in the second column. Columns are projected gas density (left) and projected SF averaged over the recent 10 Myr (middle) and over the recent 100 Myr (right).  The galaxy rotates clockwise.}
\label{fig:edgeon}
\end{figure*}

\begin{figure}
\includegraphics[trim=300 1290 250 160,clip,width=0.365\textwidth,type=png,ext=.png,read=.png]{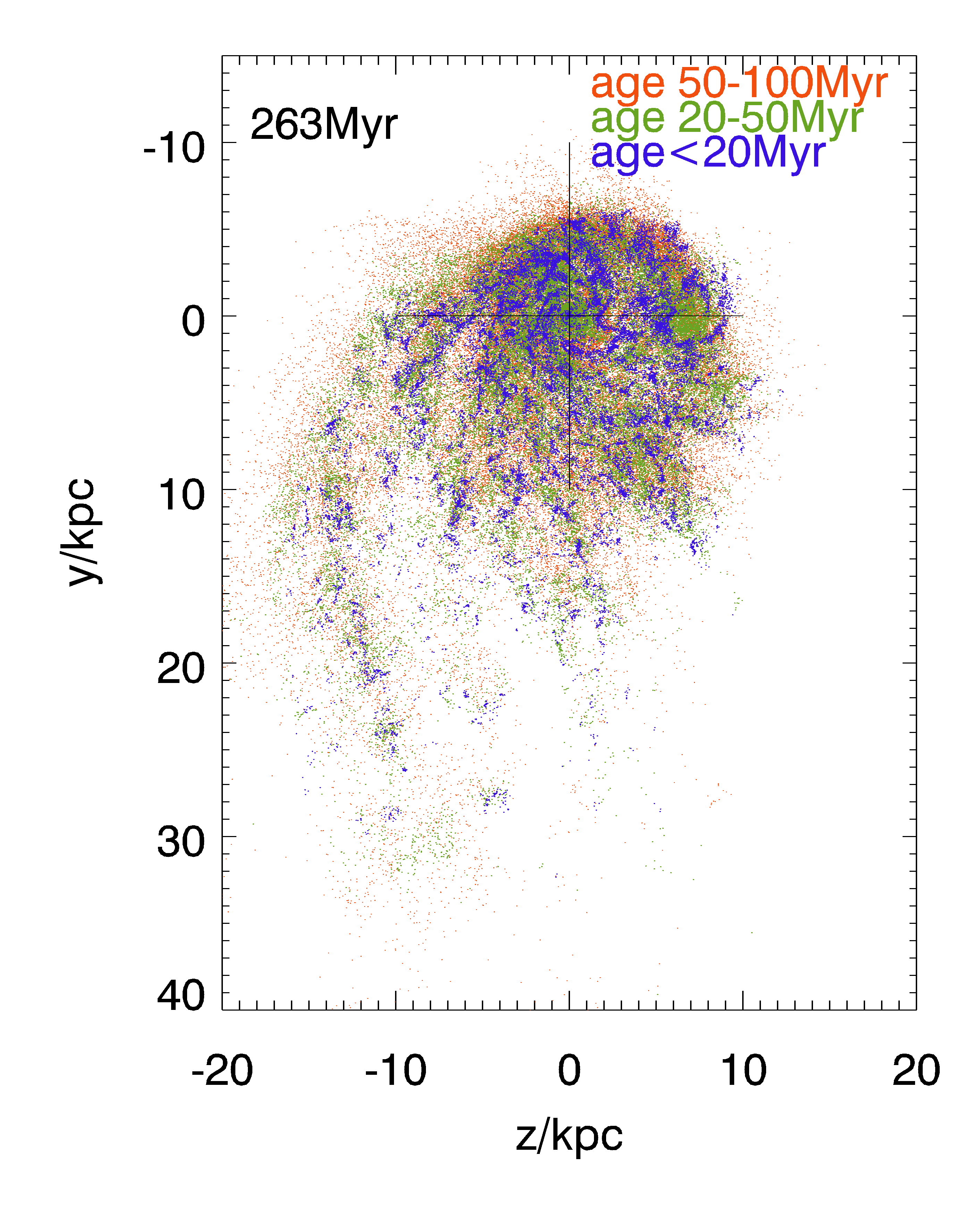}
\newline
\phantom{xxxx}
\includegraphics[trim=180 110 550 200,clip,angle=-90,origin=c,width=0.28\textwidth,type=png,ext=.png,read=.png]{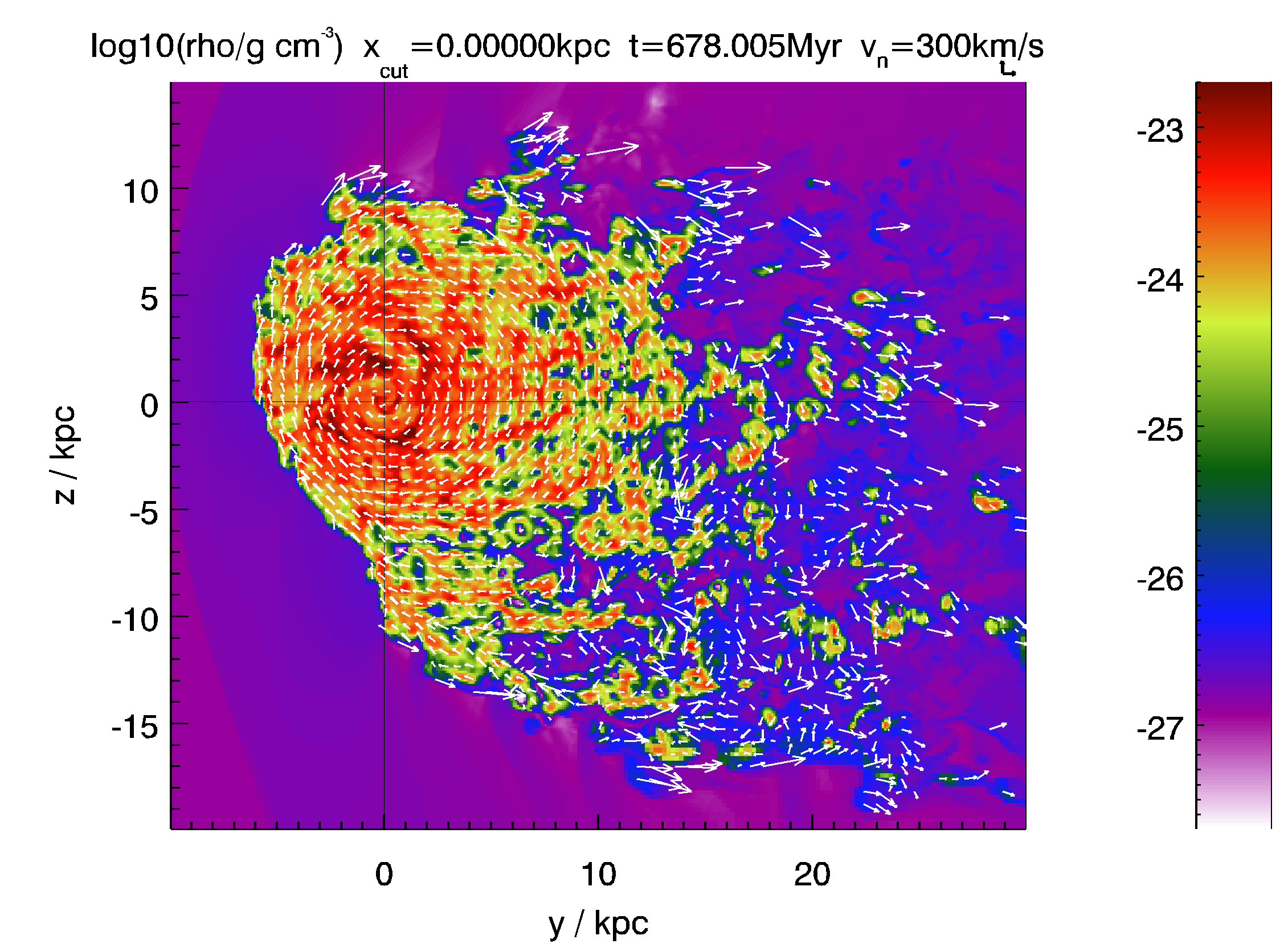}
\includegraphics[trim=1970 230 100 0,clip,angle=360,origin=c,width=0.04\textwidth,type=png,ext=.png,read=.png]{slice_dens_x_0.00000_ismflw_0339}
\caption{Top: Ages of stellar particles in edge-on stripping, last timestep of in Fig.~\ref{fig:edgeon}.  Bottom: velocity structure of the ISM in the disk plane at the same timestep (velocity vectors overlaid on  color-coded logarithmic density slice). 
\label{fig:edgeon_details}}
\end{figure}

Fig.~\ref{fig:edgeon} shows a time series of edge-on stripping, viewed face-on. The interaction between the disk rotation and the ICM determines the appearance of the stripped galaxy for a few $100\Myr$.  The overall feature is a truncated upstream gas disk and a strongly stretched downstream disk of gas and young stars. The downstream part of this highly deformed disk would naively be interpreted as tail, but all its stars and all but the lowest-density gas out to at least 20 kpc are still bound to the galaxy, and even show remains of the rotation pattern (Fig.~\ref{fig:edgeon_details}). A clear tail of ISM and young stars is formed only after 100 Myr after shock passage because also in edge-on stripping the acceleration of the gas takes time. Again, SF occurs throughout the tail or distorted disk, and stellar age gradients are absent (Fig.~\ref{fig:edgeon_details}). The youngest stars are clumped stronger along the tail filaments. 

Note the variation of the overall tail direction with time despite the constant ICM wind.  Also note the variations in asymmetry in the upstream half of the disk perpendicular to the wind direction. At earlier times the upstream half rotating with the wind has the smaller radius, at later times the situation is reversed. Asymmetries persist at least up to $250\Myr$ after shock passage. Again, SF tracers sensitive to longer timescales should find apparent SF upstream of the gas upstream edge, and broader filaments in the tail. 

In both stripping geometries, we observe significant ISM compressions by the sudden enhancement in ICM pressure only in the outer disk and only for less than 20 Myr after shock passage. The enhanced ICM thermal+ram pressure significantly affects the ISM pressure and the SFR only at radii where the ISM will be  stripped eventually. In the inner disk, the mid-plane pressure of the cold ISM is still significantly higher than even the enhanced ICM pressure, which protects the inner ISM from being stripped as well as from experiencing enhanced SF. Due to the same reason, the edge-on gas disk does not show a significantly enhanced density along its upstream rim after the first few 10 Myr. However, right after shock passage, the sudden increase of ICM pressure leads to a temporary wave of SF enhancement moving from the upstream edge along the side rims of the galaxy to the downstream edge between 10 and 25 Myr after the shock passage.  The local enhancement in SFR is about a factor 4, the global SFR is boosted only moderately by a factor of 1.5 for a short duration of $\sim 15 \Myr$. The strength of this initial burst is stronger at lower resolution because the compression region reaches further into the galaxy. However, in neither case  the inner 5 kpc are affected (about the stripping radius in the face-on case), and all quasi-equilibrium states are independent of resolution.

\section{Discussion}
%
Our simulations reproduce the overall observed morphology of stripped spirals (e.g., O12, E14, \citealt{Jachym2014}), i.e., their filamentary tails with knots of star formation up to 30 kpc length. They also explain the overall asymmetry of the disks. Our simulations  shed light on the dynamical timescales of the star-forming tails. It takes at least 10s of Myrs, possibly more, for substantial star-forming tails of $> 10\Kpc$ length to develop simply because of the inertia of the to-be-stripped gas. Similarly, an ICM shock will not be able to instantly 'turn' the direction of any pre-existing tails that the galaxies may have from their ongoing infall into their host cluster. This could explain the unusual orientation of the tail of F0083 in A2744 w.r.t.~the observed X-ray shock (O12). Our simulations predict that the jellyfish morphology should exist for galaxies overrun by an ICM shock between $\sim$ several  $10\Myr$ and a few $100\Myr$ after the shock passage. These numbers bracket the measured age of the SF in the tails of the A2744 jellyfish (O12), and match the vicinity of enhanced SF in cluster spirals observed in the 'Sausage' cluster (\citealt{Stroe2013}). 

We showed that star-forming tails can persist up to a few 100 Myr after the shock passage. With typical shock velocities of a few $1000\Kms$, galaxies as far as several 100 kpc from the current shock position may show signatures of the shock impact. A more precise assessment of the dynamical evolution of the tails requires taking into account the subsequent variation of the ICM environment of the galaxy as it moves through the cluster. 

Most of the jellyfish reported by \citet{Ebeling2014} reside in only mildly disturbed clusters, indicating that also gradual stripping can lead to a jellyfish morphology. However, most Virgo spirals show no spectacular star forming tails. The impact of gradual stripping will be the focus of a companion paper. 

The instantaneous stripping presented here does not reproduce the very straight, narrow, and parallel trails of young stars observed in ESO137-001 (Sun et al, in prep., HST photo release (\url{http://www.spacetelescope.org/news/heic1404/})) or in IC 3418 (\citealt{Hester2010,Kenney2014}), neither the  age gradients observed for the latter. We speculate that these details depend to a large degree on the stripping history, i.e., the variation of the ICM wind with time, but also draping of magnetic fields along the tail (\citealt{Ruszkowski2014}) or thermal conductivity between the ICM and the ISM could play a role. 

So far simulations  have led to contradicting conclusions about whether the increasing ambient and ram pressure can enhance SF in the remaining gas disk and tail or not. This question is  not  trivial  due to the multi-phase structure of the interstellar medium (ISM), which makes it unclear how pressure enhancements can propagate from the diffuse gas into the dense, star-forming gas. Using the SPH code GADGET \citep{Springel2003} and its hybrid star formation and stellar feedback description,  \citet{Kronberger2008} and \citet{Kapferer2009} found that, depending on wind strength, the SF in the remaining disk is boosted by factors of 2 to 10 compared to isolated galaxies, and that for stronger ram pressures a major fraction of SF can take place in the tail at distances beyond several 10s of kpc downstream of the disk. The enhancement of SF  lasted over several 100 Myr. In contrast, the grid code simulations of  TB do not find a significant enhancement of SF in the remaining gas disk, and only low levels of SF in the stripped tails. The reason for this discrepancy might be that the simulations of  TB12 and our simulations resolve higher gas densities such that SF takes place in  denser gas. We demonstrate that SF enhancements take  place only in regions of sufficiently low initial ISM pressure, which will be stripped soon afterwards. 

The jellyfish observed by O12 and E14 appear to have enhanced SF also in their disks, although quantitative  measurements have not yet been taken. Extrapolating our results to stronger ram pressures, ram pressures strong enough to eventually strip the whole galaxy should moderately enhance the SF also in the central disk prior to gas removal, which could explain the observed features.

\vspace{-0.5cm}
\section*{Acknowledgements}

E.R.~acknowledges the support of the Priority Programme 
ÓPhysics of the ISMÓ of the DFG (German Research Foundation) and  the supercomputing grant NIC 
 6006 at the John-Neumann Institut at the 
Forschungszentrum J\"ulich. M.S.O. acknowledges the funding support from the Australian Research Council through a Super Science Fellowship (ARC FS110200023). We gratefully acknowledge helpful discussions with Paul Nulsen and Ralph Kraft. 

\vspace{-0.5cm}

%
\bibliographystyle{mn2e}
\bibliography{library}

\bsp

\label{lastpage}
\end{document}